# Sunspot Observations during the Maunder Minimum from the Correspondence of John Flamsteed

V.M.S. Carrasco[1,2] • J.M. Vaquero[2,3]

[1] Departamento de Física, Universidad de Extremadura, Badajoz, Spain.

[2] Instituto Universitario de Investigación del Agua, Cambio Climático y Sostenibilidad (IACYS), Universidad de Extremadura, Badajoz, Spain

[3] Departamento de Física, Universidad de Extremadura, Mérida, Spain [e-mail: jvaquero@unex.es].

**Abstract:** We compile and analyze the sunspot observations made by John Flamsteed for the period 1672 – 1703, corresponding to the second part of the Maunder Minimum, which appear in the correspondence of this famous astronomer. We include in an appendix the original texts of the sunspot records kept by Flamsteed. We compute an estimate of the level of solar activity using these records, and compare the results with the latest reconstructions of solar activity during the Maunder Minimum, obtaining values characteristic of a grand solar minimum. Finally, we discuss a phenomenon observed and described by Stephen Gray in 1705 that has been interpreted as a white-light flare.

**Keyword:** Solar cycle, observations • Sunspots, statistics.

# 1. Introduction

The Maunder Minimum (hereafter MM) has been the only grand minimum of solar activity during the last 400 years (Eddy, 1976; Soon and Yaskell, 2003), when telescopic sunspot observations are available. It is therefore a phenomenon of great interest for astrophysicists and geoscientists (Carrasco, Villalba Álvarez, and Vaquero, 2015; Vaquero *et al.*, 2015). Recently, Zolotova and Ponyavin (2015) have hypothesized that the MM was simply a secular minimum, although Usoskin *et al.* (2015) show several errors in that study.

The Group Sunspot Number (GSN) was developed by Hoyt and Schatten (1998) (hereafter HS98). Although several articles (Vaquero, 2007; Vaquero, Trigo, and Gallego, 2012; Cliver, Clette, and Svalgaard., 2013; Clette *et al.*, 2014) have pointed to some problems in the database and method used by HS98, the effort made by those researchers was impressive. Their compilation of sunspot records preserved since the MM is an exceptional contribution for the international community. Hoyt and Schatten compiled a large number of sunspot observations during the MM, including the sunspot records by John Flamsteed, the first Astronomer Royal. In this work, we review the correspondence of Flamsteed (edited by Forbes, Murdin, and Wilmoth, 1997) to look for information about solar activity during the MM, because this historical source was not used by HS98. The letters written and received by Flamsteed during the period 1666 – 1719 were compiled in the book entitled *The correspondence of John Flamsteed, the First Astronomer Royal* (Forbes, Murdin, and Wilmoth, 1997). The correspondence is enumerated and organized into three volumes: i) volume 1, period 1666 – 1682, letters 1 – 450, ii) volume 2, period 1682 – 1703, letters 451 – 900, and iii) volume 3, period 1703 – 1719, letters 901 – 1515. Note that the dates that appear in the correspondence are the Julian calendar dates.

The aim of this article is to recover and to analyze the information about sunspots during the MM that are preserved in the correspondence of John Flamsteed. In the next section, we shall describe the historical source. In Section 3, we shall analyze the sunspot records recovered from the correspondence of Flamsteed. A critical analysis of a supposed solar flare observed by Stephen Gray in 1705 is presented in Section 4. Finally, we draw some conclusions about the level of solar activity during the MM in Section 5.

## 2. Descriptions of Sunspot Observations

John Flamsteed (1646 – 1719) is considered to be the premier star cataloguer of his time and was the first Astronomer Royal of England (Birks, 1999). Flamsteed is also considered to be one of the main observers of sunspots during the second part of the MM according to HS98 reconstruction of solar activity (Hoyt and Schatten, 1995, 1998). Moreover, Flamsteed also made measurements of the solar diameter. The main part of his sunspot observations was published in the book *Historia Coelestis Brittannica* (Flamsteed, 1725).

We have read the correspondence of Flamsteed, collecting the texts where explicit sunspot observations are described. It is important to point out that HS98 did not use this historical source, maybe because they finished the task of compiling the sunspot sources before the publication of the correspondence of Flamsteed by Forbes, Murdin and Wilmoth (1997). A part of the observations were recovered by HS98 using other historical sources (Flamsteed, 1684; Flamsteed, 1725; Wolf, 1859). In addition to the texts of the letters with information about sunspots, there are two figures in the correspondence: i) the first figure describes the sunspot of 27 July 1676, showing the positions, size, and number of sunspots recorded, and ii) the other figure contains the path that the sunspot of April – May 1684 followed on the solar disc.

In this work, we include an appendix with the texts of the correspondence of Flamsteed giving information about sunspot observations. The recovery of sunspot information from this new source is critical to fully understanding the sunspot observations made by John Flamsteed, a leading solar observer during the latter part of the MM.

3. **Analysis**

We have carefully checked the explicit sunspot records made by Flamsteed available in his correspondence, and we provide the original texts in the Appendix of this article. The Appendix contains information with the number of the letter in the compilation of Forbes, Murdin and Wilmoth (1997), the date of the letter, and the transcriptions of the relevant texts on sunspot observations for each letter. Table 1 lists the different periods in which Flamsteed explicitly mentions sunspots, indicating whether the information registered about sunspot observations is general or specific. General information is assigned when Flamsteed notes sunspot observations in a general period without explicit dates. Specific information is assigned when Flamsteed gives information about the presence or absence of sunspots for explicit days. For the period of October and December 1676, June 1684, and May and June 1703, the available information is general. In the remaining cases (January – February 1672, July and October – November 1676, April – May 1684, and June – July 1703), we have found specific information about sunspot records. The dates of these periods are Julian calendar dates.

Table 2 presents the explicit sunspot observations by Flamsteed available in his correspondence. The dates used in this list are not those of the Julian calendar, then current in England, as are presented in the correspondence (Forbes, Murdin, and Wilmoth, 1997). We have transformed these dates into the Gregorian calendar. Table 2 gives information about the presence or absence of sunspots recorded by Flamsteed. Also indicated is whether the record is "new" or is available in the HS98 database. The only information about the telescope used by Flamsteed is that he made the solar observations in 1672 with a long telescope of 164½ inches (≈ 4.2 metres) (see Appendix, Letter 87). We have recovered 43 explicit observations of sunspots made by Flamsteed in this documental source. In total, 22 observations are "new" records and 21 observations are compiled in the work of HS98.

We have calculated the total number of active and quiet days recorded by Flamsteed from sunspot observations available in his correspondence (Table 3). Comparison of the values presented in Table 3 and those in Table 1 of Hoyt and Schatten (1995) shows significant differences. First, the number of active days is lower in this work for the years 1676 and 1684. This is because some sunspot observations are not described in the correspondence by Flamsteed in those years. For the year 1703, the number of active days in this work is greater because we incorporate a new period in which Flamsteed recorded sunspots. However, the main difference between the two works lies in the number of quiet days recorded. This difference is due to the fact that this present work collects only those observations in which Flamsteed makes explicit reference to observations of sunspots. In contrast, HS98 considered all solar observations by Flamsteed to complete their database, including astrometric observations that were not specifically made to record sunspots (mainly measurements of meridian solar altitudes).

If Flamsteed did not indicate in his solar records the presence or absence of sunspots, HS98 considered that there were no spots on the solar disc.

The reconstruction of solar activity from astrometric observations of the Sun should be done with extreme caution because it might not accurately reproduce the real behavior of the Sun. For example, Vaquero and Gallego (2014) have compared some usual indices of solar activity such as sunspot number and area with solar activity as reflected in the annotations of the manuscripts of solar-meridian observations made at the Royal Observatory of the Spanish Army during 1833 – 1840. The results indicated that the information obtained about sunspots using astrometry records should either be discarded for the reconstruction of solar activity or, at least, used with extreme caution.

Furthermore, it is clear that Flamsteed recorded no sunspots in his astrometric observations when other observers did record sunspots. Flamsteed says that the spot observed during April and May 1684 was the only spot on the Sun that had appeared in the previous seven years and a half (Letter 512). From 1677 to 1683, according to the HS98 database, three days were labeled with "zero sunspots" by Flamsteed. However other observers did observe sunspots in these three days. In 1698, Flamsteed said flatly that he had not observed sunspots since 1684 (Letters 745 and 747). However, other observers did record sunspots after 1684. During the years in which Flamsteed observed the presence of sunspots (1676, 1684, and 1703), there are several days in which Flamsteed did not register spots while other astronomers did (sometimes several observers). For example, on 28 June 1684, Flamsteed registered "zero sunspots" but La Hire, Cassini, Kirch, Eimmart and Hevelius observed sunspots on the solar disc, according to HS98. Therefore, one must be wary of establishing "zero spots" in Flamsteed's observations that were not explicitly aimed at the observation of sunspots.

In order to evaluate the solar activity according to the sunspot observations recorded in Flamsteed's correspondence, we have estimated of the annual GSN using the relationship proposed by Kovaltsov, Usoskin and Mursula (2004):

$$GSN = 19 \, F_a^{1.25}$$

where $F_a$ is the annual fraction of active days. This relationship works well for low values of sunspot number (typically GSN < 30). The fraction of active days is equal to $N_a/N$, where $N_a$ is the number of active days and $N$ is the total number of days with observations (sum of the number of active and quiet days). Furthermore, the computation of the uncertainty of $N_a$ is a standard probability problem: from a box with $N$ balls (white or black), $n$ balls are taken randomly ($r$ of which are blacks). We want to know the total number of black balls contained in the box. In order to solve this problem, it is usual the use of the hypergeometric probability distribution (Kovaltsov, Usoskin, and Mursula, 2004):

$$p(s) = \frac{s!(N-s)!}{(s-r)!(N-s-n+r)!} \cdot \frac{n!(N-n)!}{(n-r)!N!r!}$$

where $N$ is the number of days in a year (365 or 366) and $s$ is the total number of active days within the year which is to be estimated. Using this distribution, we calculated the most probable value of $s$. Note that we are assuming that the observations of Flamsteed are random and independent. Clearly, this is not really true because there are a greater number of observations during days when the Sun had some spots. However, the result is that we are providing an upper limit for $F_a$ and GSN for the Flamsteed observation period during the MM.

Thus, Figure 1 shows the GSN values (black dots) for the observations recovered by Hoyt and Schatten (1995) with their mean value (grey line) calculated for the period 1676 – 1703. In addition, Figure 1 shows the average level of solar activity calculated for the period from 1672 to 1703 using the relationship proposed by Kovaltsov, Usoskin and Mursula (2004), the hypergeometric distribution, and the available sunspot observations described by Flamsteed in his correspondence (solid blue). Also, the upper limit of this solar activity with a confidence interval of 99 % for the period 1672 – 1703 is shown in Figure 1 (red-dashed line). The mean level of solar activity obtained from the observations of Flamsteed available in his correspondence (GSN ≈ 12.5) is significantly greater than that obtained in the work of Hoyt and Schatten (1995) (GSN ≈ 1). This difference is because HS98 assigned "zero" values to all the dates when Flamsteed reported an observation but said nothing about sunspots. However, the upper limit of solar activity using only the explicit information of sunspots is still only approximately 15.5, implying that solar activity during the period 1672 to 1703, the second half of the MM, was very low. These values are compatible with a grand minimum of solar activity.

**4.     Was a White-Light Flare Observed by Stephen Gray in 1705?**

Hoyt and Schatten (1996) indicated that Stephen Gray of Canterbury recorded a "flash of lightning" near a sunspot on 27 December 1705. They interpreted this observation as a white-light flare, and hence as evidence of the capacity of solar observers during the Maunder Minimum. Note that the first observation of a white-light flare in the scientific literature until this forgotten report is the famous "1859 flare of Carrington" (Neidig and Cliver, 1983).

This information is reproduced in *The correspondence of John Flamsteed, the First Astronomer Royal*. In a letter sent by Stephen Gray to Flamsteed, the observation of a new phenomenon is described. The original text is:

Letter 1062 (1705 December 27): […] I am in Persute [pursuit] of a new Phenomenon of the suns Spots [sunspots.] I say Persute [pursuit] because though I suspect that I have seen it more than once yet I have often looked for it without success tis [sic] this there seems sometimes to Proceed from the West side the Spot as it were flash of lightening which moves round the spot by the north to the East and is there extinguished generaly [generally] but sometimes it arives [arrives] to the south before extinction this is soon after followed by an other [another] such like Phenomenon they succeed each other in about a second of time. the Tremulation of the Atmosphear [atmosphere] I cannot think to be the cause of this Phenomenon but however shall suspend my judgment till I have confirmed the apearance [appearance] by more observations. […]

From this description by Stephen Gray, we think that the nature of this phenomenon is unclear. On the one hand, Gray thought he had observed this phenomenon in the past. But it is unlikely to frequently observe a phenomenon such as a white-light flare. On the other hand, Gray points out that hopes to confirm the phenomenon from more observations. However, Gray did not report another similar phenomenon again. In any case, although solar flares can brighten briefly, one second is an extremely short period of time for a solar flare to develop. We note this observation was carried out approximately 150 years before the event observed by Carrington in 1859 (Cliver and Keer, 2012), considered as the first white-light flare observed. If the phenomenon observed by Gray is a solar flare, it would be the first record of this kind in history. We think that Gray's record does not describe a solar flare although it is hard to know whether he saw a white light flare or not.

## 5. Conclusions

In this work, we have recovered and analyzed sunspot observations made by Flamsteed during the MM (from 1672 to 1703) in the correspondence of John Flamsteed, a source not consulted by HS98. We have included an appendix with the original texts of Flamsteed's correspondence that explicitly indicate information about sunspots. We have calculated the number of active and quiet days. Thus, we have obtained an average value and an upper limit of the solar activity. Our values significantly exceed the average values obtained by Hoyt and Schatten (1995). This can be explained by the large number of observations with zero values present in the database of HS98. Most of these zero values included in the HS98 database come from astrometric observations of the Sun. The upper limit of solar activity obtained in this work (GSN ≈ 15) indicates that during the period 1672 – 1703 solar activity was compatible with a grand minimum of solar activity. This result contradicts the qualitative reconstruction of solar activity obtained recently by Zolotova and Ponyavin (2015). Finally, we have discussed the original text by Stephen Gray about a possible observation of a white-light flare in 1705. We show that the description by Gray is clearly not consistent with the description of a typical solar flare. Finally, we want to emphasize that a revision of the solar activity during the MM needs to be carried out taking into account the original historical observations.

**Acknowledgements**

Support from the FEDER-Junta de Extremadura (Research Group Grants 15137) and from the Ministerio de Economía y Competitividad of the Spanish Government (AYA2014-57556-P) is gratefully acknowledged.

**Disclosure of Potential Conflicts of Interest**

The authors declare that they have no conflicts of interest

**Table 1.** Periods of the explicit sunspot records from Flamsteed's correspondence and type of information (general or specific) for each period. The periods are Julian calendar dates.

| PERIOD | INFORMATION |
| --- | --- |
| January – February 1672 | Specific |
| July 1676 | Specific |
| October 1676 | General |
| October – November 1676 | Specific |
| December 1676 | General |
| April – May 1684 | Specific |
| June 1684 | General |
| May 1703 | General |
| June 1703 | General |
| June – July 1703 | Specific |

**Table 2.** Daily explicit sunspot observations by Flamsteed according to his correspondence. The first three columns contain the year, month, and day of the observation (Gregorian calendar). Information about the presence or absence of sunspots is given in the fourth column. The fifth column shows whether the record is contained in the HS98 database.

| Year | Month | Day | Description | New or old record |
|------|-------|-----|-------------|-------------------|
| 1672 | Jan | 15 | NO SUNSPOT | N |
|      | Jan | 19 | NO SUNSPOT | N |
|      | Jan | 20 | NO SUNSPOT | N |
|      | Jan | 21 | NO SUNSPOT | N |
|      | Jan | 22 | NO SUNSPOT | N |
|      | Jan | 25 | NO SUNSPOT | N |
|      | Feb | 11 | NO SUNSPOT | N |
| 1676 | Aug | 6  | SUNSPOT    | O |
|      | Nov | 3  | NO SUNSPOT | N |
|      | Nov | 4  | NO SUNSPOT | N |
|      | Nov | 19 | SUNSPOT    | O |
|      | Nov | 20 | SUNSPOT    | N |
|      | Nov | 21 | SUNSPOT    | N |
|      | Nov | 22 | SUNSPOT    | O |
|      | Nov | 23 | SUNSPOT    | N |
|      | Nov | 24 | SUNSPOT    | O |
|  | Nov | 25 | SUNSPOT | O |
|  | Nov | 26 | SUNSPOT | O |
|  | Nov | 27 | SUNSPOT | N |
|  | Nov | 28 | SUNSPOT | N |
|  | Nov | 29 | SUNSPOT | O |
| 1684 | Apr | 20 | NO SUNSPOT | N |
|  | May | 5 | SUNSPOT | O |
|  | May | 6 | SUNSPOT | O |
|  | May | 7 | SUNSPOT | O |
|  | May | 8 | SUNSPOT | O |
|  | May | 9 | SUNSPOT | O |
|  | May | 10 | SUNSPOT | O |
|  | May | 11 | SUNSPOT | O |
|  | May | 12 | SUNSPOT | O |
|  | May | 13 | SUNSPOT | O |
|  | May | 14 | SUNSPOT | O |
|  | May | 15 | SUNSPOT | O |
|  | May | 16 | SUNSPOT | O |
|  | May | 17 | SUNSPOT | O |
|  | May | 18 | NO SUNSPOT | O |
| 1703 | Jul | 8 | SUNSPOT | N |
|  | Jul | 9 | SUNSPOT | N |
|  | Jul | 10 | SUNSPOT | N |
|  | Jul | 11 | SUNSPOT | N |
|  | Jul | 12 | SUNSPOT | N |

| | | | |
|---|---|---|---|
| Jul | 13 | SUNSPOT | N |
| Jul | 16 | NO SUNSPOT | N |

**Table 3.** Number of active, quiet, and total days according to the explicit sunspot observations available in Flamsteed's correspondence.

| Year | Active | Quiet | Total |
|------|--------|-------|-------|
| 1672 | 0      | 7     | 7     |
| 1676 | 12     | 2     | 14    |
| 1684 | 13     | 2     | 15    |
| 1703 | 6      | 1     | 7     |

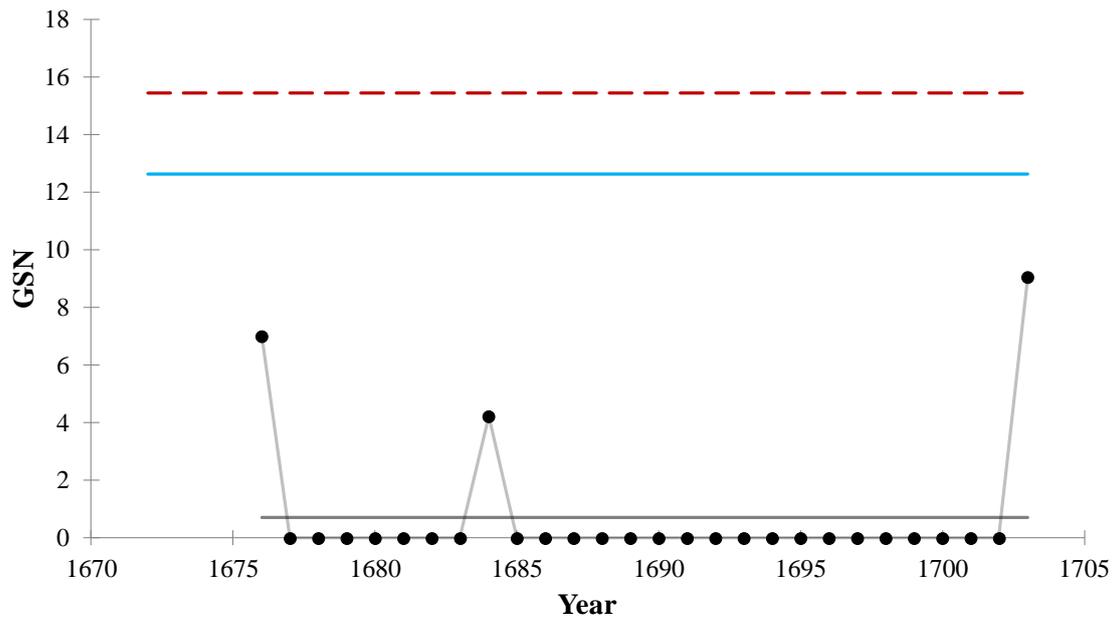

**Figure 1.** Group Sunspot Number (black dots) calculated from solar observations made by John Flamsteed according to Hoyt and Schatten (1995) and the average value (grey line) for the period 1676 – 1703. Average group sunspot number (solid blue line) and upper limit with a confidence interval of 99 % (dashed-red line) obtained in this work

**Appendix. Sunspot Observations Made by John Flamsteed Recovered from his Correspondence.**

We present here the transcriptions of the relevant texts on sunspot observations by Flamsteed. Also, we have also indicated the number of the letter that is assigned in the compilation by Forbes, Murdin and Wilmoth (1997) and the date of the letter (in Julian calendar).

**Transcription**

Letter 84: 31 January 1671/2, Flamsteed to Collins

[…] all the times I have veiwed the sun I could never see any Macula upon him but his whole disck. perfectly cleare. […]

Letter 85: 5 February 1671/2, Flamsteed to Oldenburg

[…] Thursday last being February 1st instant was very cleare so that veiwing the sun severall times I found his body cleare from spots. […]

Letter 87: 10 February 1671/2, Flamsteed to Collins

[…] all these diameters were observed in my tube of $164^1/_2$ inches, at none of these times [January 5, 9, 10, 11, 12, 15 and February 1] could I find any spot under the sun but hee has beene constantly cleare. […]

Letter 268: 27 July 1676, Flamsteed to Moore

Honored Sir

Whilest I was takeing some altitudes of the sun this morneing to correct the times of my last nights observations, I found upon his face a considerable large spot. which because it was the firs I ever saw, and a thinge you have not yet beene acquanted with I thought it might be no unwelcome novelty to enforme you of: the notes I tooke of it were these:

hor. corr.

|  h ' " | | |
|---|---|---|
| 9.51.25 | Macula a limbo Solis proximo. | 710=5 – 43 ML |
| 9.55 | distantia Az: limbi Solis dextri et Mac. | |
| | | 1380=11 – 21 MZ |
| 10.03.45 | eodem distantia | 1383=11 - 22 MZ |

These notes I dare not affirme to be very praecise, because the wind sometimes shooke the tube, and groweing stronger would not permit me to measure the ☉s diameter which therefore I have derived from my former observations. 31'. 46"

hence the distance of the spot from the ☉s Centr.     10.10 ☉M

from the verticall passeing by it                                  4.31 MA

The parallactick Angle at $10^h.03¾'$ by calculation I find

$$43°.55½':$$

Hence the spot in Consequence of the Suns center     9'.34''☉e

with south latitude from it                                  3.25½ e.M

I have drawne the figure of this spot and its position as it appeared through the tube, inverted: hold but the bottom of the paper upwards and you have the true appearance with its due position in respect of the ecliptick and verticall then passeing by the suns center.

The diameter of the spot in its broadest place was equall to ♃s or 50''. its length not wholly double. about 1⅓'.

It seemed a little cloven in the middle and had two thin cloudy spots following it like those in the figure.

It has not yet measured over above ⅓ part of its way through the sun. so that I suppose wee may see it yet 8 days if it breake not and dissipate before it have finished its jorney over his face which I much suspect, by reason that it seemed to part in its middle. I am apt to thinke the small spots following it were but parts of it broken off from it, and that therefore it was much larger whilest on the other side of the sun: [...]

Letter 274: 11 December 1676, Flamsteed to Towneley

[...] I can salve the severall appearances of the spot that appeared in the sun in October to the 24 or 25. Returned againe the 9 of November and after a revolution past came againe upon him the 6th Instant where you will find it till the 18th or 19th. [...]

Letter 450: 29 May 1682, Flamsteed to Molyneux

[...] As for spots in the sun I have never seene more then two the first in August 1676 which was large but broke into peeces and almost disappeared before it had passed through the visible diske of the sun. another in October November and December following I observed which was more Compact and made three revolutions before it was dissolved yet there was no large one. [...]

Letter 512: 2 May 1684, Flamsteed to Molyneux

[…] I tell you that this day was sevennight being the 25 of. Aprill in the morneing as I was takeing the distance of ☿ from the sun I discovered a large spot entred a little within the following limbe of his diske, the time of theire semirevolutions is 13 dayes and more then an halfe: but on the 10th of April at Noone I observed his Meridional distance from the vertex it was then cleare scarce 15 dayes before and I am confident there was then no spot on his face so that this certeinely had its rise in his latent hemisphere tis neare 7½ yeares since I saw one before they have of late beene so scarce how ever frequent in the days of Galileo and Scheiner By the next dayes observations I stated its longitude in the suns diske from his aequinoctiall Colure and

from thence determined in what pointes of his face it would be visible till it passed of his limbe into his opposite superficies these I give you in the Included figure: On the 8th of May in the Morneing it passes out of his diske, and if it have consistence enough to hold a second revolution it will be seene entred his following limbe againe on the 22 describeing a line very neare streight in its passage over him. The Theory of the spots is briefly delivered in my preface to the Doctrine of the Sphere inserted into Sir Jonas Moores workes, where I suppose the revolution of any point in the sun *ad fixas* to be compleate in 25 dayes six hours praecise. […]

### Letter 519: 8 July 1684, Flamsteed to Bernard

[…] The magnitude and consistency of the spot emerging from the Sun seems to me to be such that I believe it may still last for another solar rotation. If it does, it will appear again visibly on the limb of the Sun's disc on the 13th of this present July, and it will be seen inside it, [moving] towards the following limb, on the 14th. I think that this one spot will not last for three solar rotations, but that two or more have arisen, having spewed themselves forth in the vicinity of the first; or rather, if indeed you would see [how] I am brought to my opinion, that Etna-like mountains have been raised up from the thick subcutaneous matter of the Sun. For during the second revolution I saw two quite large spots almost two minutes apart, with rather pale companions in the second revolution. In the shape of these, observed in the middle of the Sun, neither the first or the newest that I observed can in any way be re-established. […]

### Letter 745: 3 May 1698, Flamsteed to Leigh

[…] As for Spots in the Sun there have been none since the Year 1684. you may acquaint Mr. Ayres of it and that which is published in the forreigne prints is a Romance. the sun haveing been as clear of late yeares as ever, and I have seldom omitted Observeing him at Noon when it was clear. […]

Letter 747: 19 May 1698, Flamsteed to Leigh

[…] I told you in my last no spots have been seen in the Sun since 1684 all the storys you have heard of them are a Scilly Romance Spread by such as call themselves witty men to abuse the Credulous and not to be heeded J F. […]

Letter 905: 3 July 1703, Flamsteed to Sharp

I returne an imediate Answer to yours of the 29th past because this week since Monday last I have seene spots in Sun. which tho they are no novelty to me may be so to you they are advanced a little beyond the middle of the sun so that if this letter meets with a speedy conveyance you may find them before they turne out of him tho they change their shape dayly which makes me thinke they are shallow and will scarce continue another revolution.

Wee have seene of them ever since the middle of May and in June one of them returned that was a pretty dense one I expec[t] to see it within his antecedent limbe againe this day or tomorrow. [...]

Letter 906: 8 July 1703, Flamsteed to Lister

[…] Wee have seen great variety of spots in the sun. since May last. On tuesday last he was cleare and had none. but I expect a return of some this day. there is nothing to be learnt by them more then we know already and therefore. I should not have mentiond this but that since the year 1684 to the present I have seen none on him.